\documentclass[aps,prl,preprint,groupaddress]{revtex4}
\usepackage{graphicx}
\usepackage{dcolumn}
\usepackage{bm}
\usepackage{color}
\usepackage{amsmath}
\begin{document}
%
\title{Negative-pressure-induced helimagnetism in ferromagnetic cubic perovskites Sr$_{1-x}$Ba$_{x}$CoO$_{3}$}
%
\author{H. Sakai$^{1,2,3}$}\email[Corresponding author: ]{sakai@phys.sci.osaka-u.ac.jp}
\author{S. Yokoyama$^{1}$}
\author{A. Kuwabara$^{1}$}
\author{J. S. White$^{4}$}
\author{E. Can\'evet$^{4,5}$}
\author{H. M. R\o nnow$^{6,7}$}
\author{T. Koretsune$^{8}$}
\author{R. Arita$^{1,7}$}
\author{A. Miyake$^{9}$}
\author{M. Tokunaga$^{9}$}
\author{Y. Tokura$^{1,7}$}
\author{S. Ishiwata$^{1,3}$}\email[Corresponding author: ]{ishiwata@ap.t.u-tokyo.ac.jp}
\affiliation{$^1$Department of Applied Physics, University of Tokyo, Tokyo 113-8656, Japan.\\
$^2$Department of Physics, Osaka University, Toyonaka, Osaka 560-0043, Japan.\\
$^3$PRESTO, Japan Science and Technology Agency, Kawaguchi, Saitama 332-0012, Japan.\\
$^4$Laboratory for Neutron Scattering and Imaging, Paul Scherrer Institut, Villigen CH-5232, Switzerland.\\
$^5$Department of Physics, Technical University of Denmark, 2800 Kgs. Lyngby, Denmark.\\
$^6$Laboratory for Quantum Magnetism, \'Ecole Polytechnique F\'ed\'erale de Lausanne (EPFL), Lausanne CH-1015, Switzerland.\\
$^7$RIKEN Center for Emergent Matter Science (CEMS), Wako, 351-0198, Japan.\\
$^8$Department of Physics, Tohoku University, Sendai, 980-8578, Japan.\\
$^9$The Institute for Solid State Physics, University of Tokyo, Kashiwa 277-8581, Japan}
%
\begin{abstract}
Helimagnetic materials are identified as promising for novel spintronic applications.
Since helical spin order is manifested as a compromise of competing magnetic exchange interactions, its emergence is limited by unique constraints imposed by the crystalline lattice and the interaction geometries, as exemplified by the multiferroic perovskite manganites with large orthorhombic distortion.
Here we show that a simple cubic perovskite SrCoO$_3$ with room-temperature ferromagnetism has the potential to host helimagnetic order upon isotropic lattice expansion.
Increasing the Ba content $x$ in Sr$_{1-x}$Ba$_x$CoO$_3$ continuously expands the cubic lattice, eventually suppressing the ferromagnetic order near $x$=0.4, where helimagnetic correlations are observed as incommensurate diffuse magnetic scattering by neutron diffraction measurements.
The emergence of helimagnetism is semi-quantitatively reproduced by first-principles calculations, leading to the conjecture that a simple cubic lattice with strong $d$-$p$ hybridisation can exhibit a variety of novel magnetic phases originating from competing exchange interactions.
\end{abstract}
%
%
\maketitle
%
Exploration of helimagnetic materials and the relevant novel quantum phases, such as multiferroic\cite{Cheong2007NatureMat,Tokura2010AdvancedMater} and skyrmion lattice\cite{Nagaosa2013NatNanotech} phases, has been one of the hottest research topics in the field of condensed matter physics as well as spintronics.
For helimagnetic-based spintronics, perovskite-type transition-metal oxides can be promising candidates when finely controlling the interplay between spin, orbital, and lattice degrees of freedom.
This is exemplified by the spin-spiral-driven ferroelectricity in perovskite-type manganites showing giant magnetoelectric effects\cite{Kimura2003Nature,Kenzelmann2005PRL}.
The spin-spiral orders in these manganites arise from the frustration of the exchange interactions between the Mn$^{3+}$ spins, which is enhanced by the orthorhombic (GdFeO$_3$-type) distortion with orbital order\cite{Kimura2003PRB}. 
On the other hand, perovskite-type ferrites $A$FeO$_3$ ($A$=Sr, Ba) with unusually high-valence ions of Fe$^{4+}$ show helical spin order even in a simple cubic lattice \cite{Oda1977JPSJ,Reehuis2012PRB,Ishiwata2011PRB,Hayashi2011Angew,Hayashi2013JPSJ}.
Recently, itinerant helimagnets with high lattice symmetry have been predicted to host a rich variety of topologically nontrivial magnetic structures typified by the skyrmion lattice\cite{Okubo2012PRL,Azhar2017PRL,Hayami2017PRB}.
Thus, cubic perovskite $A$FeO$_3$ generates strong interest as a promising class of materials for novel topological magnets\cite{Ishiwata2011PRB,Ishiwata2018unpub}.
%
\par
%
The origin of the helimagnetic order in SrFeO$_3$ is ascribable to a strong hybridisation between the Fe 3$d$ and O 2$p$ orbitals with the negative charge transfer energy $\Delta$ yielding itinerant ligand holes\cite{Bocquet1992PRB,Mostovoy2005PRL,Azhar2017PRL}.
Reflecting the correlation between the nature of the magnetic order and the $p$-$d$ hybridisation, the magnetic structure of $A$FeO$_3$ is strongly dependent on the size of the $A$-site ion, since this affects the bandwidth of the system.
While SrFeO$_3$ and BaFeO$_3$ share the same simple cubic structure, they show a different type of helical spin order\cite{Hayashi2011Angew,Hayashi2013JPSJ}; the propagation vector of BaFeO$_3$ is aligned with the $<$0 0 1$>$ direction, whereas that of SrFeO$_3$ is aligned along the $<$1 1 1$>$ direction.
%
\par
%
Perovskite-type $A$CoO$_3$ ($A$=Ca, Sr) with the high valence state of Co$^{4+}$ is another interesting example of a negative-$\Delta$ system having itinerant ligand holes\cite{Bezdicka1993Zanorg,Abbate2002PRB,Balamurugan2006PRB,Long2011JPhysC,Xia2017PRM,Osaka2017PRB}.
SrCoO$_3$ has a cubic structure and shows metallic and ferromagnetic behaviour at ambient pressure with the ferromagnetic transition temperature $T_{\rm C}$=305 K.
Very recently, CaCoO$_3$ was reported to be an antiferromagnetic metal with a distorted perovskite structure\cite{Osaka2017PRB}.
Thus, the size of the $A$-site ions emerges as a dominant factor for the magnetic order in $A$CoO$_3$ as well as for $A$FeO$_3$.
Furthermore, unlike the ferrites, cobaltates display a variability of the spin states (high/intermediate/low) \cite{Potze1995PRB, Saitoh1997PRB, Asai1998JPSJ}, which emerges as another factor influencing the magnetic order.
In this work, we explore novel magnetic order in the perovskite-type cobaltates with negative $\Delta$, and study the effect of lattice expansion on the ferromagnetic order in SrCoO$_3$ by synthesising single-crystalline samples of Sr$_{1-x}$Ba$_{x}$CoO$_{3}$ using a high-pressure technique\cite{Sakai2010PRB,Ishiwata2011PRB,Long2011JPhysC,Long2012PRB,SM}.
By increasing the Ba content, the ferromagnetic transition is suppressed monotonically and then disappears around $x\!\sim\!0.4$, yet all the while keeping the cubic lattice structure and intermediate spin state.
Neutron diffraction studies reveal that the compound with $x$=0.4 hosts short-range helimagnetic correlations described by an incommensurate propagation vector along the $<$1 1 1$>$ direction at the lowest temperature.
We shall discuss the origin of the emergent helimagnetic instability in Sr$_{1-x}$Ba$_{x}$CoO$_{3}$ on the basis of first-principles calculations.
%
%
%
\begin{figure}
\begin{center}
\includegraphics[width=.7\linewidth]{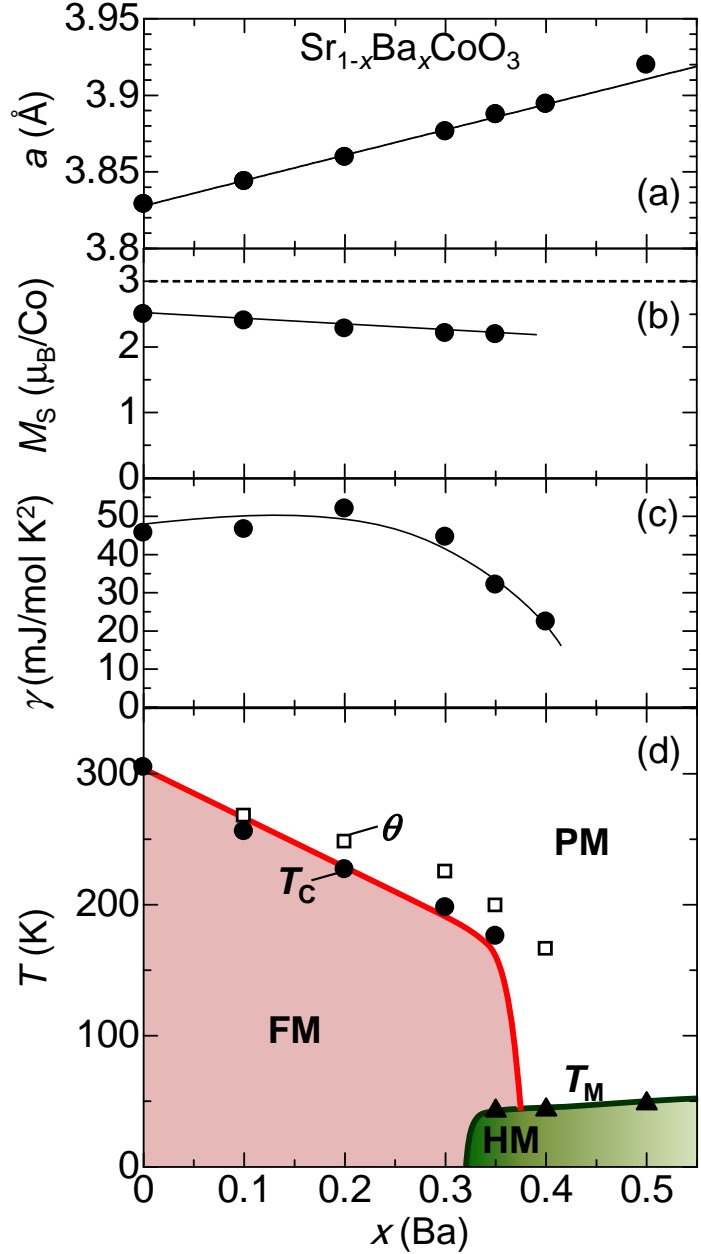}
\end{center}
\caption{\label{fig:phase}(color online) $x$ dependence of the (a) cubic lattice constant $a$ at room temperature, (b) saturation magnetisation $M_{\rm S}$ at the lowest temperature (2-4 K), and (c) electronic specific heat coefficient $\gamma$ for Sr$_{1-x}$Ba$_{x}$CoO$_{3}$ ($0\!\le\! x\!\le\!0.4$) single crystals. (d) Magnetic phase diagram as functions of temperature $T$ and $x$, where the ferromagnetic (FM) and helimagnetic (HM) transition temperatures are denoted by closed circles ($T_{\rm C}$) and closed triangles ($T_{\rm M}$), respectively. The Weiss temperatures $\theta$ are also plotted as open squares. The data for $x$=0 are from Refs. \onlinecite{Balamurugan2006PRB,Long2011JPhysC}.}
\end{figure}
%
\par
%
The room temperature structure was determined from x-ray diffraction on powder obtained by grinding pieces of single crystals.
For all $x$, the diffraction profiles showed the cubic symmetry ($Pm\bar{3}m$) with the $x$-dependent lattice constant $a$ as depicted in Fig. 1(a).
The value of $a$ increases linearly with $x$ from 0 to 0.4 in accordance with Vegard's law, whereas that for $x$ = 0.5 slightly deviates upward from the straight line fit for the data in the range of $x\!\le\! 0.4$.
This slight increase in $a$ for $x$ = 0.5 may arise due to inevitable oxygen release after the high-pressure treatment, which is furthermore consistent with thermogravimetry measurements revealing a slight increase in the oxygen deficiency (see Fig. S1).
We note that Sr$_{1-x}$Ba$_{x}$MnO$_{3}$ hosting a similar tolerance factor exhibits a structural transition from cubic to tetragonal symmetry for $x\!\ge\!0.45$\cite{Sakai2011PRL}, while Sr$_{1-x}$Ba$_{x}$CoO$_{3}$ remains cubic up to $x$=0.5.
%
\par
%
Figure 2(a) shows the temperature dependence of the magnetisation $M$ at 0.01 T for Sr$_{1-x}$Ba$_{x}$CoO$_{3}$, measured upon heating after a field cooling process.
With increasing $x$ from 0.1 to 0.35, the ferromagnetic Curie temperature $T_{\rm C}$ [defined as the inflection point of $M(T)$] decreases systematically from 256 K to 176 K. 
Accordingly, the Weiss temperature $\theta$, determined by fitting the Curie-Weiss law to data above 250 K (Fig. S2), also decreases in value comparable to $T_{\rm C}$.
For $x$=0.35, the increase in $M$ below $T_{\rm C}$ is largely reduced, and instead a clear drop in $M$ upon cooling manifests itself at $T_{\rm M}$ [$\sim$43 K, defined as the peak position in $dM(T)/dT$].
For $x$ = 0.4, the ferromagnetic transition disappears and the drop in $M$ is more conspicuous.
As clearly seen in Fig. 2(b), the value of $M$ falls steeply below $T_{\rm M}$ in both measurements after the field and zero field cooling runs.
This behaviour indicates that the compound with $x$=0.4 has an antiferromagnetic-like ground state, though the observed positive value of $\theta\!\sim\! 170$ K [Fig. 1(d)] implies that this state is unlikely to be a simple antiferromagnetic.
Indeed, as shown later, it turns out to host helimagnetic correlations (see Fig. 4).
For $x$=0.5, although an anomaly in $M$ is discernible around 50 K, the steep drop in the field cooling run seen for $x$=0.4 changes to a cusp-like anomaly, suggesting a spin-glass-like transition.
Figure 1(d) summarises the overall magnetic phase diagram of Sr$_{1-x}$Ba$_{x}$CoO$_{3}$ as functions of temperature and $x$, which reveals that the competition between ferromagnetism and helimagnetism is sensitively dependent on the size of the cubic lattice, i.e., the Co-O bond length.
In what follows, we shall show the $x$ dependence of various physical properties for the present system. 
%
\par
%
To check the spin state of Co$^{4+}$ ions upon Ba substitution, we have estimated the saturation magnetisation $M_{\rm S}$ by measuring the field profile of $M$ at the lowest temperature (2$-$4 K)[Fig. 2(c)].
For $x$=0.1$-$0.3 with the ferromagnetic ground state, the $M$ value is saturated above $\sim$2 T.
However, the saturation field is substantially enhanced as $x$ increases to 0.35, where the helimagnetic instability sets in.
Although the value of $M$ barely saturates at above $\sim$30 T for $x$=0.35, the magnetisation does not reach $M_{\rm S}$ even under magnetic fields up to 40 T for $x\!\ge\!0.4$,.
Figure 1(b) summarises the value of $M_{\rm S}$ as a function of $x$.
In spite of a significant decrease in $T_{\rm C}$, the value of $M_{\rm S}$ is almost constant ($\sim$2.2$-$2.5$\mu_B$/Co) with $x$ up to 0.35.
The spin state thus appears to remain in a nearly intermediate, i.e., $S\!=\!3/2$, configuration\cite{Long2011JPhysC}, which is likely to be strongly hybridised with the ligand hole states [see the inset to Fig. 2(c)].
%
\begin{figure}
\begin{center}
\includegraphics[width=0.8\linewidth]{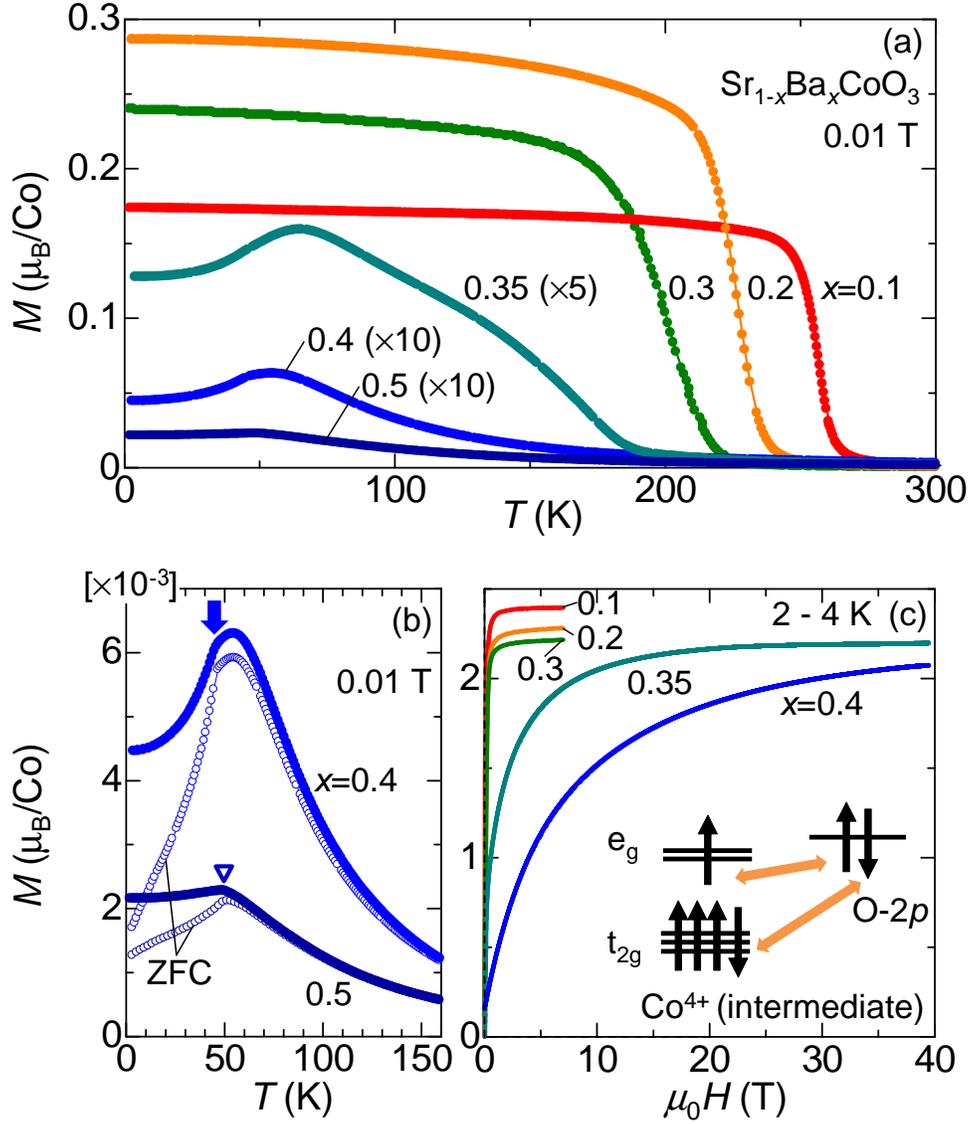}
\end{center}
\caption{\label{fig:MT_MH}(color online) (a) Temperature ($T$) profiles of magnetisation ($M$) at 0.01 T for Sr$_{1-x}$Ba$_{x}$CoO$_{3}$ ($0.1\!\le\! x\!\le\!0.4$) single crystals. (b) Low-$T$ profiles of $M$ at 0.01 T for $x$=0.4 and 0.5, where open circles denote the data in a warming run after zero-field cooling. (c) Field dependence of $M$ at the lowest temperature (2-4 K) for $x$=0.1-0.4. The profiles up to $\sim$40 T were measured with a pulsed magnet. Inset shows schematic diagram of intermediate spin configuration of Co$^{4+}$. The $d^6\underline{L}$ configuration ($\underline{L}$ denotes an oxygen ligand hole) is likely to be stabilised by strong hybridisation between the Co 3$d$ and O 2$p$ orbitals.} 
\end{figure}
%
\par
%
The temperature dependence of the resistivity for the Sr$_{1-x}$Ba$_{x}$CoO$_{3}$ single crystals is shown in Fig. 3(a).
The $x$=0.1 and 0.2 crystals show clear metallic behaviour with resistivity lower than 1 m$\Omega$cm similarly to SrCoO$_3$\cite{Long2011JPhysC}, and a weak resistivity anomaly is discernible around the ferromagnetic transition as denoted by closed triangles.
With increasing $x$, the resistivity increases at almost all temperatures and becomes less temperature-dependent.
The former indicates a cross-over towards a high-resistivity state, which is consistent with a reduction in density of states at the Fermi energy for $x\!\ge\! 0.35$ signified by the specific heat data [Fig. 3(b)].
On the other hand, the latter may result from scattering due to spin-fluctuations persisting at low temperatures, which become enhanced by the competition between the ferromagnetic and helimagnetic correlations.
In addition, significant disorder effects due to both of the Sr/Ba solid solution and oxygen vacancies may also play a role on such diffusive transport with high resistivity.
%
\par
%
Results of specific heat measurements shown in Figs. 3(b) and 1(c) suggest a significant change in density of states at the Fermi energy upon Ba substitution.
For $x$=0.1$-$0.3 with a ferromagnetic ground state, the temperature profile of $C$ is reasonably well described by $C/T=\gamma+\beta T^{2}$, where $\gamma$ is the electronic specific-heat coefficient [Fig. 3(b)]\cite{note1}.
The large $\gamma$ values of 45$-$50 mJ/mol K$^2$ for $x$=0.1$-$0.3 imply a moderately strong electron correlation, which is comparable to the $\gamma$ value of SrCoO$_3$ reported previously\cite{Balamurugan2006PRB}.
For $x\!\ge\! 0.35$, the $C/T$ value plotted as a function of $T^2$ shows a clear downturn toward the lowest temperature, indicating a substantial decrease in the density of states at the Fermi energy.
The $\gamma$ values estimated from $C/T$ are plotted as a function of $x$ in Fig. 1(c).
Although the $\gamma$ value is more or less constant up to $x$=0.3, it steeply decreases for $x\!\ge\!0.35$, indicating the density of states to become reduced in the helimagnetic state.
The $\gamma$ value for $x$=0.4 ($\sim$ 20 mJ/mol K$^2$) is less than half those for $x$=0$-$0.3.
Thus, a partial energy gap might be formed around the Fermi energy in the helimagnetic phase, being consistent with the large increase in resistivity for $x\!\ge\! 0.35$.
Note here that in spite of the large variation in the density of states upon $x$, the $M_{\rm S}$ value remains almost independent of $x$, which implies that the $M_{\rm S}$ value is mainly determined by the localised $t_{2g}$ majority spins.
In fact, according to a recent dynamical mean field theory, it can be expected that the itinerant electronic states at the Fermi energy consist of both $e_{\rm g}$ majority and $t_{\rm 2g}$ minority spin states\cite{Kunes2012PRL}.
Since these two states have opposite spin states, they may provide a small contribution to $M_{\rm S}$.
%
\begin{figure}
\begin{center}
\includegraphics[width=.8\linewidth]{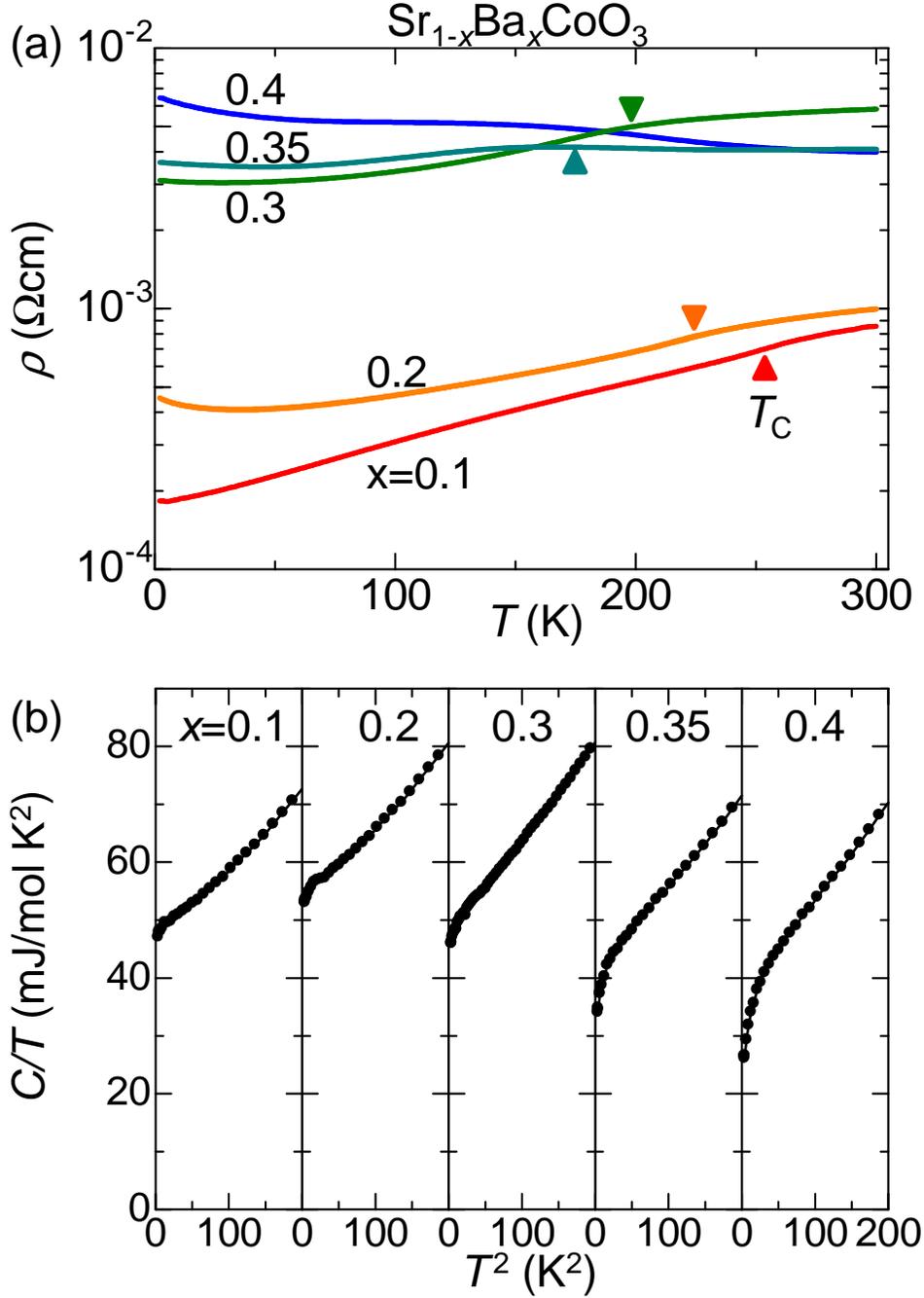}
\end{center}
\caption{\label{fig:RT_CT}(color online) (a) Temperature profile of resistivity ($\rho$) at zero field for Sr$_{1-x}$Ba$_{x}$CoO$_{3}$ ($0.1\!\le\!x\!\le\!0.4$) single crystals. Closed triangles denote the ferromagnetic transition temperature $T_{\rm C}$. (b) Specific heat divided by temperature ($C/T$) versus $T^{2}$ for $0.1\!\le\!x\!\le\!0.4$.}
\end{figure}
%
\par
%
Next we show results of neutron diffraction measurements of the magnetic correlations for $x$=0.4.
Figures 4(a) to 4(c) show maps of the neutron diffraction intensity around the (0 0 1) Bragg reflection for a single-crystalline sample at selected temperatures.
The data in the panels (a) and (b) are plotted after subtraction of the data at 150 K in the paramagnetic state, which visualises the magnetic scattering intensity around the (0 0 1) reflection.
At 50 K just above $T_{\rm M}$ ($\sim$44 K), a nearly isotropic distribution of magnetic diffuse scattering is seen, which implies the presence of ferromagnetic correlations.
At 1.5 K below $T_{\rm M}$, part of the magnetic scattering is observed to be anisotropically distributed, forming lobe shapes along the $<$1 1 1$>$ directions.
The anisotropic distribution of the magnetic intensity that onsets below $T_{\rm M}$ can be identified by subtracting the neutron diffraction intensity map at 50 K from that at 1.5 K as shown in Fig. 4(c).
From this we find that incommensurate magnetic peaks are discernible at four equivalent positions along the $<$1 1 1$>$ directions.
The presence of the incommensurate peaks is also clear from the line-cut profile of the intensity along $<$1 1 1$>$ [Fig. 4(d)], with no intensity observed in the profile along the $<$0 0 1$>$ direction [Fig. 4(e)].
From a Gaussian fit to the line-cut profile, the peak position is determined to be ${\bf\it Q}=(0\ 0\  1) \pm (\delta\ \delta\ \delta)$ with $\delta=0.079(5)$, and most likely originates from incommensurately modulated helimagnetic correlations with a periodicity of approximately 28 \AA.
From the half-width at half maximum (HWHM) of the peaks the magnetic correlation length is estimated to be 12 \AA.
Such short-range correlation might reflect strong fluctuations induced by the presence of the competing ferromagnetic instability, which is consistent with the fact that $x$=0.4 is located near the ferromagnetic-helimagnetic phase boundary.
In addition to this, disorder effects arising from the oxygen vacancies and solid solution may contribute to the short range nature of the correlations.
We have thus revealed that the ferromagnetic ground state in Sr$_{1-x}$Ba$_{x}$CoO$_{3}$ tends towards incommensurate helimagnetism as $x$ increases from 0 to 0.4.
%
\begin{figure}
\begin{center}
\includegraphics[width=.9\linewidth]{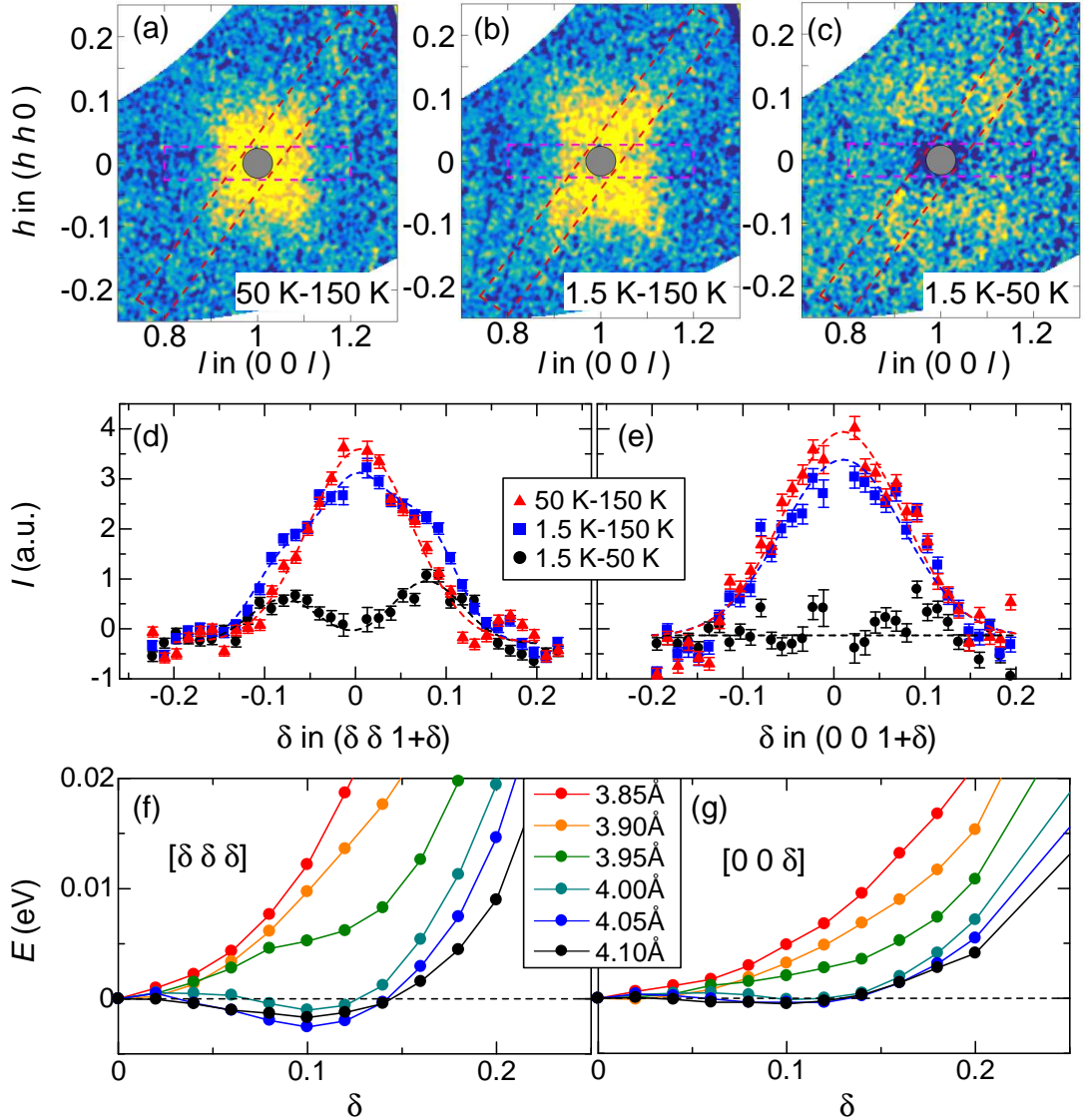}
\end{center}
\caption{\label{fig:Neutron}(color online) Reciprocal-space maps of neutron diffraction intensities for $x$=0.4 around the (0 0 1) reflection in the ($h$ $h$ $l$) plane measured at (a) 50 K and at (b, c) 1.5 K. To clearly show the magnetic scattering intensity, the intensities in panels (a), (b), and (c) are plotted after the subtraction of those at 150 K, 150 K, and 50 K, respectively. Line profiles of the intensity in panels (a) to (c), for cuts along the (d) $<$1 1 1$>$ and (e) $<$0 0 1$>$ directions. The cuts along the $<$1 1 1$>$ ($<$0 0 1$>$) directions were evaluated using the red (pink) boxes indicated in panels (a) to (c). Calculated total energy as a function of magnetic propagation vector (f) [$\delta$\ $\delta$\ $\delta$] and (g) [0\ 0\ $\delta$] for various lattice constants (\AA), where the energy is plotted with respect to the energy of the FM state $E(\delta=0)$.}
\end{figure}
%
\par
%
The instability towards helimagnetic order in Sr$_{1-x}$Ba$_{x}$CoO$_{3}$ can also be reproduced by first-principles calculations.
Figures 4(f) and 4(g) show the $\delta$ dependence of the total energy for various lattice constants, and for the magnetic propagation vectors of [$\delta$\ $\delta$\ $\delta$] and [0\ 0\ $\delta$], respectively.
By increasing the lattice constant, the local minimum energy at $\delta\!\sim\! 0.1$ becomes global minimum for [$\delta$, $\delta$, $\delta$], indicating a first-order ferromagnetic-to-helimagnetic transition.
Since the minimum energy for [$\delta$\ $\delta$\ $\delta$] is lower than that for [0\ 0\ $\delta$], the propagation vector of the most stable helimagnetic order is determined to be [$\delta$\ $\delta$\ $\delta$] with $\delta\!\sim\! 0.1$, which is close to the experimental result.
The origin of the helimagnetic order may be attributed to a competition between the ferromagnetic double exchange interaction and the antiferromagnetic superexchange interaction\cite{Li2012PRB85,Li2012PRB86}, i.e., the former becomes weakened compared with the latter as the lattice expands.
Such a tendency is in fact expected according to first-principles calculations done for the perovskite-type ferrites that takes into consideration a change in spin moment\cite{Li2012PRB86}.
Alternatively, the hybridisation of the oxygen $2p$ orbital relevant to the negative $\Delta$ may play an important role by modulating the nature of double exchange interaction\cite{Mostovoy2005PRL,Azhar2017PRL}.
It remains a future theoretical work to reveal the detailed variation in exchange interactions and the mechanism of the helimagnetic transition upon lattice expansion.
%
\par
%
In combination with first-principles calculations, this work demonstrates that a simple cubic Co$^{4+}$-O lattice has the potential to host incommensurate helical spin order that competes with the ferromagnetic order.
The point here is that the stability of these phases in Sr$_{1-x}$Ba$_x$CoO$_3$ can be controlled by the lattice size, i.e., the bandwidth.
Therefore, it can be anticipated that the Sr$_{1-x}$Ba$_x$CoO$_3$ thin film heterostructures serve as a novel helimagnetic-based spintronics, where the ferromagnetic and helimagnetic states are selectively stabilised by the epitaxial strain.
%
\begin{acknowledgements}
The authors are grateful to M. Mostovoy for fruitful discussions.
Neutron scattering was performed at the Swiss Neutron Source SINQ, Paul Scherrer Instutute (PSI).
 The work was in part supported by the JST PRESTO Hyper-nano-space design toward Innovative Functionality (Grant No. JPMJPR1412), the JSPS Grant-in-Aid for Scientific Research (A) No. 17H01195, the Asahi Glass Foundation, the Swiss National Science Foundation through grants 153451 and 166298 and the Sinergia network on Nanoskyrmionics (grant CRSII5-171003), and the European Research Council project CONQUEST. E.C. acknowledges support from the Danish Research Council for Science and Nature through DANSCATT.
\end{acknowledgements}
%

%

\begin{thebibliography}{}
\bibitem{note1} A slight deviation from the straight line might arise from the magnon contribution in the ferromagnetically ordered phase. The $\gamma$ value was estimated by extrapolating the $C/T$ data close to the lowest temperature.
\end{thebibliography}
\end{document}